\documentclass[useAMS,usenatbib]{mn2e}

\usepackage{epsfig}
\usepackage{amsmath}
\usepackage{amssymb}
\usepackage{natbib}
\usepackage{threeparttable} 
\usepackage[hyphens]{url}

\usepackage{rotating}
\usepackage{graphicx}

\usepackage{epstopdf}

\usepackage{booktabs,array}

\usepackage{color}

\newcommand{\rxs}{1RXS J171824.2--402934}
\newcommand{\rxh}{1RXH J173523.7--354013}
\newcommand{\xte}{XTE J1719--291}

\newcommand{\xtecool}{XTE J1709--267}
\newcommand{\xtejoel}{XTE J1701--462}
\newcommand{\igr}{IGR~J17494--3030}

\def\apj{ApJ}
\def\mnras{MNRAS}
\def\aap{A\&A}
\def\apjl{ApJL}

\def\ssr{Space Sci. Rev.}

\def \mnras {MNRAS}
\def \apj {ApJ}

\def \apjl {ApJL}
\def \aap {A\&A}

\def \atel {ATel}

\def\na{New A}

\def\lesssim{\mathrel{\hbox{\rlap{\hbox{\lower4pt\hbox{$\sim$}}}\hbox{$<$}}}}
\let\la=\lesssim
\def\gtrsim{\mathrel{\hbox{\rlap{\hbox{\lower4pt\hbox{$\sim$}}}\hbox{$>$}}}}
\let\ga=\gtrsim

\def\arcsec{\hbox{$^{\prime\prime}$}}

\def\sol{~\mathrm{M}_\odot}
\def\lx{L_\mathrm{X}}
\def\lxpeak{$L_\mathrm{X}^{peak}$}

\def\Nh{N$_{H}$}

\def\chired{$\chi^{2}_{\nu}$}
\def\chis{$\chi^{2}$}

\newcommand{\chan}{\textit{Chandra}}
\newcommand{\swift}{\textit{Swift}}
\newcommand{\xmm}{\textit{XMM-Newton}}

\newcommand{\inte}{\textit{INTEGRAL}}

\newcommand{\kpc}{\mathrm{kpc}}
\newcommand{\lum}{\mathrm{erg~s}^{-1}}
\newcommand{\flux}{\mathrm{erg~cm}^{-2}~\mathrm{s}^{-1}}
\newcommand{\cnts}{\mathrm{counts~s}^{-1}}
\newcommand{\nh}{\mathrm{cm}^{-2}}

\title[The very-faint X-ray transient \igr]{\xmm\ and \swift\ spectroscopy of the newly discovered very-faint X-ray transient \igr}
\author[M. Armas Padilla et al.]
{M. Armas Padilla$^{1}$\thanks{e-mail: m.armaspadilla@uva.nl}, 
R. Wijnands$^{1}$, N. Degenaar$^{2}$\thanks{Hubble fellow}\\
$^{1}$Astronomical Institute ``Anton Pannekoek", 
University of Amsterdam, 
Postbus 94249, 1090 GE Amsterdam, The Netherlands\\
$^{2}$Department of Astronomy,
University of Michigan,
500 Church Street, Ann Arbor, MI 48109, USA\\
}

\begin{document}

\date{DRAFT VERSION}

\pagerange{\pageref{firstpage}--\pageref{lastpage}} \pubyear{0000}

\maketitle

\label{firstpage}

\vspace{-0.2cm}
\begin{abstract} A growing group of low-mass X-ray binaries are found
to be accreting at very-faint X-ray luminosities of $< 10^{36}$ erg
s$^{-1}$ (2--10 keV). Once such system is the new X-ray transient \igr.
We present \swift\ and \xmm\ observations obtained during its 2012
discovery outburst. The Swift observations trace the peak of the
outburst, which reached a luminosity of $\sim7\times 10^{35}$ (D/8
kpc)$^2$ erg s$^{-1}$ (2--10 keV). The XMM-Newton data were obtained
when the outburst had decayed to an intensity of $\sim8\times10^{34}$ (D/8
kpc)$^2$ erg s$^{-1}$. The spectrum can be described by a power-law
with an index of $\Gamma \sim1.7$ and requires an additional soft component
with a blackbody temperature of $\sim$0.37~keV (contributing $\sim$20\% to the
total unabsorbed flux in the 0.5--10 keV band). Given the similarities
with high-quality spectra of very-faint neutron star low-mass X-ray
binaries, we suggest that the compact primary in \igr\ is a neutron
star. Interestingly, the source intensity decreased rapidly during the
$\sim$12 hr \xmm\ observation, which was accompanied by a
decrease in inferred temperature. We interpret the soft spectral
component as arising from the neutron star surface due to low-level
accretion, and propose that the observed decline in intensity was the
result of a decrease in the mass-accretion rate onto the neutron star.
\end{abstract}



\section{Introduction}\label{sec:intro}

Low mass X-ray binaries (LMXBs) are composed of 
a black hole or a neutron star that accretes material from a (sub) solar
companion star that overflows its Roche lobe. A large fraction of
LMXBs remain in a dim quiescence state, during which no or
very little accretion occurs and the X-ray luminosity ($\lx$) is
$\sim10^{30-33}\lum$. However, once in a while, the sources
experience outburst events in which their accretion rate increases
drastically and consequently their X-ray brightness as well. LMXBs can
be classified depending on the maximum 2--10 keV luminosity (\lxpeak) that they reach. Those
systems that can obtain \lxpeak of $\sim10^{37-39}\lum$ are called
\textit{(very--)bright} systems. The \textit{faint} ones can reach
\lxpeak $\sim10^{36-37}\lum$, and the sources that display \lxpeak of
only $\sim10^{34-36}\lum$ are called \textit{very faint} X-ray
binaries \citep[VFXBs; ][]{Wijnands2006}. Despite significant progress over the last few years in our
understanding of the behavior of those VFXBs \citep{Muno2005,campana2009,Degenaar2009,degenaar2010,ArmasPadilla2011,armaspadilla2013,ArmasPadilla2013b}, much remains unclear about
them (e.g., the mechanism behind their
low luminosities is not understood). 

One recently discovered VFXB is \igr.
The source was first detected during the Galactic centre \inte\
observation performed on March 17--19, 2012
\citep{Boissay2012}.  Multiple \swift\ X-ray telescope (XRT) observations were obtained \citep{Bozzo2012} but $\sim$27 days after the source was 
first detected, it could not be detected anymore using 
\chan\ \citep{Chakrabarty2013}. Inspecting the near--infrared (NIR) images acquired on July 2010 as part of the NIR VVV survey
\citep{Minniti2010}, five sources could be identified within the
\swift/XRT error circle. Those sources are potential candidates
for the quiescent NIR counterpart of \igr\ although it cannot be
excluded that none of those sources is associated with the source. In this {\em Letter} we present the analysis of four \swift\
observations and our \xmm\ observation  of \igr\  obtained during the outburst
decay phase.

\vspace{-0.6cm}
\section{Observations and reduction}\label{sec:Obs}

\vspace{-0.1cm}

\begin{table*}
\caption{\swift\ and \xmm\ observation log.}
\begin{threeparttable}
\begin{tabular}{l c c c c c c}
\hline \hline
Satellite/Instrument & Mode  & Observation ID & Date & Exposure time & Count rate & Net count rate$^{a}$\\
&  &  & (yyyy-mm-dd) & (ks) & (counts s$^{-1}$) & (counts s$^{-1}$) \\
\hline
\swift/XRT & PC & 00032318001 & 2012-03-20 & 1.0 &  0.93 &0.35\\ 
\swift/XRT & WT & 00032318002 & 2012-03-23  & 1.0 &  1.89 & 1.63\\ 
\swift/XRT & PC & 00032318003 & 2012-03-26 & 0.6 &  0.80 & 0.39\\ 
\swift/XRT & PC & 00032318004 & 2012-03-30 & 1.1 &  0.28 & 0.28\\
\xmm/EPIC &  & 0694040201 & 2012-03-31 & 43.9 &  \\
\multicolumn{1}{r}{(MOS1)} & Imaging &  &  &  & 0.44 & 0.38 \\
\multicolumn{1}{r}{(MOS2)} & Imaging &  &  &  & 0.44 & 0.38 \\
\multicolumn{1}{r}{(PN)} & Timing &  &  &  &  1.34 & 0.98\\

\hline
\end{tabular}
\label{tab:obs}
\begin{tablenotes}
\item[a]{Count rate after background correction, using the annulus regions to mitigate the pile-up in \swift\ obs 00032318001 and 00032318003, and after excluding a short episode of background flaring observed during the \xmm\ observation.}
\end{tablenotes}
\end{threeparttable}
\end{table*}

\subsection{\xmm\ data}\label{subsec:xmm}
\vspace{-0.1cm}
\igr\ was observed with \xmm\  \citep{Jansen2001} on
2012 March 31 for $\sim43$~ks (see Tab.~\ref{tab:obs}). The European
Photon Imaging Camera (EPIC) on board of \xmm\ consists of two MOS
cameras \citep{Turner2001} and one PN camera \citep{Struder2001}. They
were operated in imaging (full frame window) and timing mode,
respectively.  We reduced the data and obtained scientific products
using the Science Analysis Software (SAS, v. 13.0).

We filtered out an episode of background flaring by only selecting data
for which the high-energy count rate was $<$0.25 counts s$^{-1} $
($>10$ keV) for the MOS cameras and $<$0.2 counts s$^{-1}$ (10--12
keV) for the PN. The total resulting live time is $\sim42$~ksec for
the MOS cameras and $\sim39$~ksec for the PN detector. For the MOS
cameras, we extracted the source event file using a circular region
centred on the source position and with a radius of $\sim47\arcsec$;
the background was extracted using a circular region with a radius of
$\sim107\arcsec$ centred on a source-free region. The MOS source count
rate (Tab.~\ref{tab:obs}) is below the 0.7~$\cnts$ pile-up
threshold\footnote{See Table 3 in Section 3.2.2 of the \xmm\ Handbook: \url{http://heasarc.gsfc.nasa.gov/docs/xmm/uhb/epicmode.html}}. After
correcting the PN data for the transfer inefficiency that
affects data obtained using the timing modes\footnote{See the
calibration technical note XMM-SOC-CALTN-0083}, we extracted the
source and background events by selecting the data with RAWX columns
[32:42] and [5:12], respectively. The PN source count rate (Tab.~\ref{tab:obs}) is much lower that the pile--up limit of
800~$\cnts$ for the PN timing mode$^{1}$. We generated the spectra and
the light curves as well as the response files applying the standard analysis threads. We grouped the
spectra to contain a minimum of 25 photons
per bin and rebinned the data to not to oversample the intrinsic energy resolution by a factor larger than 3.

\subsection{\swift\ data}\label{subsec:swift}

A total of 4 observations were obtained of \igr\ with the XRT
(\citealt{Burrows2005}) on board \swift\ \citep{Gehrels2004}. Three
observations were performed in photon counting (PC) mode and one in
windowed timing (WT) mode (Table \ref{tab:obs}). We processed the
data making use of the {\ttfamily HEASoft} v.6.12 software. The data
reduction was carried out running the {\ttfamily xrtpipeline}
task. For every observation, spectra, lightcurves and images were
obtained using {\ttfamily Xselect}. For the WT data we used a
circular region radius of $\sim76\arcsec$ centred on the source
position to extract the source events, and a similar region far enough
from the source to extract the
background events.  For the PC data, we used a
circular region of $\sim47\arcsec$ centred on the source position to
extract the source events, and three circular regions of similar size
for the background data. Observations 00032318001 and 00032318003 are
affected by pile--up. To mitigate the pile--up effects, we
excluded the inner $\sim11\arcsec$ (observation 00032318001) and
$\sim9\arcsec$ (00032318003) from the central part of the source
regions.  We created exposure maps and ancillary response files
following the standard \swift\ analysis
threads and we
acquired the last version of the response matrix files from the
\textsc{HEASARC} calibration data base (v.14). We grouped the spectra to have at least 20 counts per bin.

\section{Analysis and Results}\label{sec:results}

\subsection{Spectral Analysis}\label{subsec:spect}

%
\begin{table*}
\caption{Results from the spectral fits.}
\begin{threeparttable}
\begin{tabular}{l c c c c  c c c c}
\hline \hline
Instrument/ & $N_{\mathrm{H}}$ & $\Gamma$ & $kT$& $T_{\mathrm{fr}}$ & $F_{\mathrm{X, abs}}$ & $F_{\mathrm{X,unabs}}$ & $L_{\mathrm{X}}$ & $\chi^2_{\nu}$ (dof) \\
Obs ID & ($10^{22}$ $\nh$) &  & (keV) & (\%) &  \multicolumn{2}{c}{($10^{-11}$ $\flux$)}  & ($10^{35}$ $\lum$)  & \\  
\hline
\swift/XRT & \multicolumn{8}{c}{phabs*(powerlaw)}\\

00032318001 & 1.8 (fix) & $1.8 \pm 0.2$ &  -- & -- & $8.6 \pm 0.7$ &  $14.0\pm 0.2$& $10.7 \pm 0.8$ & 0.99 (15) \\ 
00032318002 & 1.8 (fix) & $1.9 \pm 0.1$ &  -- & -- & $10.2 \pm 0.4$ &  $17.4 \pm 0.1$& $13.4 \pm 0.8$ & 0.98 (79) \\ 
00032318003 & 1.8 (fix) & $1.8 \pm 0.3$ &  -- & -- & $6.9 \pm 0.9$ &  $11.3 \pm 0.3$& $8.7 \pm 0.8$ & 0.53 (9) \\ 
00032318004 & 1.8 (fix) & $2.0 \pm 0.3$ &  -- & -- & $1.8 \pm 0.2$ &  $3.4 \pm 0.2$& $2.6 \pm 0.8$ & 0.92 (13) \\ 
\hline 
\textit{XMM}/EPIC & \multicolumn{8}{c}{phabs*(powerlaw+bbodyrad)}\\

0694040201 & $1.80 \pm 0.07$ & $1.76 \pm 0.08$ &  $0.37 \pm 0.03 $ & 17.3 & $0.649 \pm 0.005$ &  $1.13 \pm 0.05$& $0.86 \pm 0.05$ & 0.94 (419) \\

 & \multicolumn{8}{c}{phabs*(powerlaw+nsatmos)}\\

	 & $1.89 \pm 0.03$ & $1.74 \pm 0.07$ &  $0.182 \pm 0.004 $ & 22.9 & $0.649 \pm 0.003 $ &  $1.18 \pm0.01 $& $0.90 \pm 0.01$ & 0.94 (420) \\

\hline
\end{tabular}
\label{tab:spec}
\begin{tablenotes}
\item[]Note. -- Quoted errors represent 90\% confidence levels. The fifth column reflects the fractional contribution of the thermal component to the total unabsorbed 0.5--10 keV flux. $F_{\mathrm{X, abs}}$ and $F_{\mathrm{X, unabs}}$ represent the absorbed and unabsorbed fluxes (0.5--10 keV), respectively. The luminosity $L_{\mathrm{X}}$ (0.5--10 keV) was calculated adopting a distance of 8~kpc. The $kT$ for the NSATMOS model is for an observer in infinity.
\end{tablenotes}
\end{threeparttable}
\end{table*}

\begin{figure}$\phantom{!h}$
\begin{center}
\includegraphics[angle=0,width=8cm]{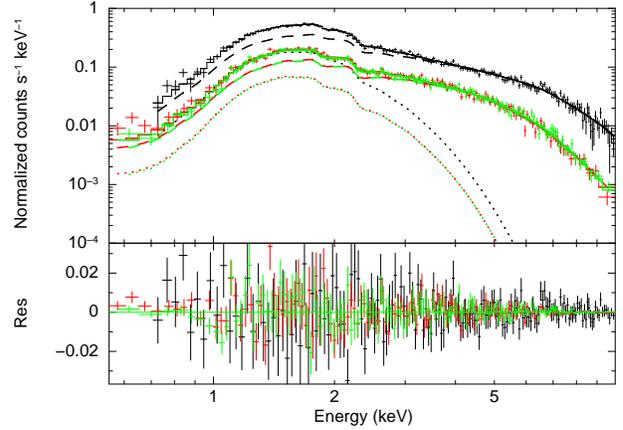}\\
\caption{PN (black), MOS1 (red) and MOS2 (green) spectra of \igr. The solid line represented the best fit using a \textsc{bbodyrad+powerlaw} model. The doted lines are the contribution of the thermal component and the dashed lines the contribution of the power law. The sub--pannel shows the residuals. }
\label{fig:spec}
\end{center}
\end{figure}

We used \textsc{XSpec} \citep[v.12.8;][]{xspec} to fit the spectra. We incorporated in our models the photoelectric
absorption component (\textsc{phabs}) to account for the interstellar
absorption. We simultaneously fitted the spectra
obtained with the different \xmm/EPIC detectors; the two MOS spectra
were fitted in the energy range 0.5-10 keV and the PN spectrum in the
0.7--10 keV range. We tied all the parameters between the three
detectors and introduced a constant factor (\textsc{constant}) to
account for cross calibration uncertainties between the
instruments. It was fixed to one for the PN spectrum and allowed to
vary freely for the MOS ones. A single power--law (\textsc{powerlaw})
model returns an hydrogen column density (\Nh) of
$1.87\times10^{22}\nh$ and a power--law photon index ($\Gamma$) of
2.14. However, with a reduced \chis (\chired) of 1.28 for 421 degrees
of freedom (dof) the fit is not acceptable. Adding a soft blackbody
component (\textsc{bbodyrad}) to the power-law model improves the
fit. The resulting spectra with the best fit model are shown in
Figure~\ref{fig:spec}.

This two components model adequately fits the spectra (\chired=0.94
for 419 dof), yielding \Nh=$1.8\times10^{22}\nh$, $\Gamma$=1.76 and
kT=0.37~keV (blackbody temperature). The inferred 0.5--10~keV
unabsorbed flux is $1.13\times10^{-11}\flux$, of which $\sim$17\% is
due to the thermal component. The full results of the fits are
summarized in Table~\ref{tab:spec}, in which the parameter errors are
given with 90\% of confidence level and the flux errors were
calculated as described by \citet{Wijnands2004}. Although the source
distance is unknown, the fact that the source position is close to the
Galactic center (\textit{l}=359.086, \textit{b}=--01.511), we assume a distance of
8~kpc for \igr\, which is consistent with the relatively high \Nh\ we obtained in the spectral fits.

The spectra were equally well fitted using neutron star atmosphere
models for the soft component. We used the NSATMOS model
\citep[][]{Heinke2007}. In this we assume that the accretor in the
system is a neutron star but we note that the nature of the primary is
still unknown. We fixed the mass of the putative neutron star to
$1.4\sol$ and the radius to 10~km. We used a distance D of 8~kpc and
we set the normalisation to 1 (i.e., the whole surface is assumed to
emit radiation), which leaves the neutron star temperature as the only
parameter free for the NSATMOS model. The obtained temperature (for an
observer at infinity) is $kT^{\infty}=0.182\pm0.004$~keV and the soft
component contributes $\sim$22\% to the total 0.5-10~keV unabsorbed
flux. The other fit parameters are, within the errors, consistent with
the values obtained when using the \textsc{bbodyrad} model (see
Table~\ref{tab:spec}).


\begin{figure}$\phantom{!h}$
\begin{center}
\includegraphics[angle=0,width=\columnwidth]{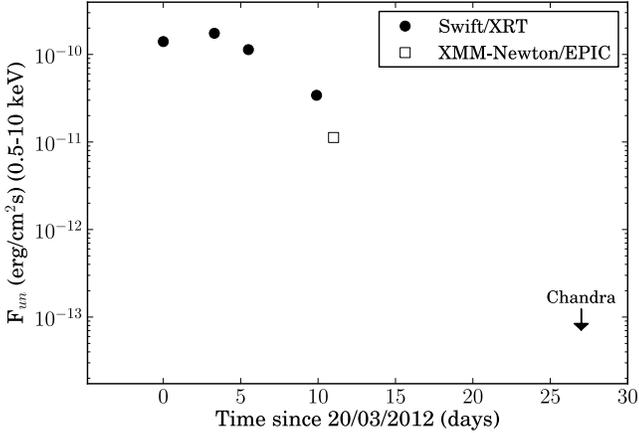}\\
\caption{The unabsorbed flux (0.5--10 keV) evolution. Black circles represent the \swift\ observations while the white square represent the \xmm\ one. The arrow is the upper limit calculated from the upper limit on the source count rate observed using \chan\ \citep{Chakrabarty2013}.  }
\label{fig:curve}
\end{center}
\end{figure}

All \swift\ spectra are well described with a single absorbed
power--law model (see Table~\ref{tab:spec}). We fixed \Nh\ to the
value obtained from the \xmm\ fit ($1.87\times10^{22}\nh$). The photon
index value is, within the errors, consistent in all observations. The
peak unabsorbed flux is $\sim1.7\times10^{-10}\flux$ (as observed
during the second \swift\ observation) which corresponds to a $\lx$ of
$\sim1.3\times10^{36}(D/8~\kpc)^{2}\lum$ (0.5--10 keV). The lowest
detected flux was observed during the \xmm\ observation. The
unabsorbed flux decreased approximately one order of magnitude in
$\sim$~8~days (see also Fig.~\ref{fig:curve}), which gives a
luminosity seen during the \xmm\ observation of
$\sim8.6\times10^{34}(D/8~\kpc)^{2}\lum$. We calculated the
0.5--10~keV unabsorbed flux upper limit from the \chan\ count rate
upper limit reported in \citet{Chakrabarty2013} using WebPIMMS. We assumed an
absorbed power--law model with \Nh=$1.87\times10^{22}\nh$ and
$\Gamma$=2.14 (the values obtained from the \xmm\ spectra). The
calculated 0.5--10~keV unabsorbed flux is $<6.8\times10^{-14}
\flux$ which results in a luminosity of
$<5.2\times10^{32}(D/8~\kpc)^{2} \lum$ (see Fig.~\ref{fig:curve}).

%

\subsection{Light curve Analysis}\label{subsec:lc}


\begin{figure}$\phantom{!h}$
\begin{center}

\includegraphics[width=\columnwidth]{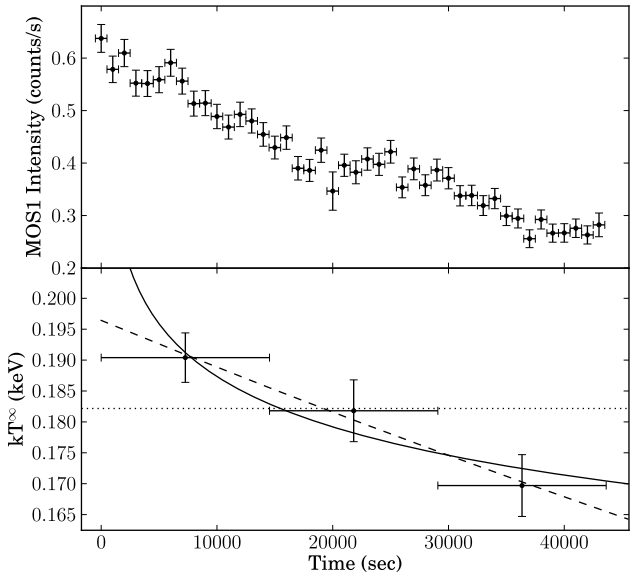}\\

\caption{\textit{Upper pannel}: the MOS1 count rate curve using a bin size of 1000 s. \textit{Bottom pannel:} The evolution in the temperature during the \xmm\ observation. The solid-line is the fit using a power-law decay model (with an index of --0.06$\pm$0.02), the dashed-line is the fit assuming an exponential decay function (with an e-folding time of $\sim$3 days), and the dotted line is the fit assuming a constant value (which does not provide an adequate fit).}
\label{fig:temp}
\end{center}
\end{figure}


The count rate curve obtained during the \xmm\ shows that the source
intensity decreased during the observation, from $\sim0.65~\cnts$ to
$\sim0.25~\cnts$ within $\sim$12 hours (see Fig.~\ref{fig:curve}). In
order to study the spectral evolution during the observation, we
divided the observation in three segments of $\sim14.5$~ksec each and
individually fitted each spectrum (both the PN and MOS spectra
simultaneously). We used the absorbed \textsc{NSATMOS + POWERLAW}
model with the \Nh\ fixed to the value obtained from the fit using the
data obtained from the whole observation (see
Sec.~\ref{subsec:spect}). While the power-law index is constant within
the errors ($\Gamma\sim1.7$), the temperature decreases from
$kT^{\infty}=0.190\pm0.004$~keV to $kT^{\infty}=0.170\pm0.005$~keV
(Fig.~\ref{fig:temp}; the errors are for 1$\sigma$). We can fit
this decrease adequately with an exponential decay function with an
e-folding time of 3.0$^{+1.4}_{-0.7}$ days or with a power-law
decay function with an index of --0.06$\pm$0.02 (see
Fig.~\ref{fig:temp})\footnote{As reference time we used the
start of the \xmm\ observation, but we found that the e-folding time
or index is not very sensitive to the exact value of the reference
time.}. We also tried a constant value, but that did not
provide an adequate fit (with a $\chi^2$ of 10.5 and a dof of 2,
resulting in a p-value of 0.005).  Both the flux observed from the
soft component as well as the power-law component decrease during the
observation. The thermal contribution to the 0.5-10 keV remained
approximately constant during the decay.

\vspace{-0.3cm}
\section{Discussion}\label{sec:Discuss}

We present the spectral analysis of the \swift\ and \xmm\ observations
of \igr\ during its 2012 outburst. If we assume a distance of 8~kpc,
the inferred peak luminosity is $\lx\sim1.3\times10^{36}\lum$ in the
energy range 0.5--10 keV. However, the luminosity in the 2--10 keV
energy range is only $\lx\sim7.7\times10^{35}\lum$, making this source
a very-faint X-ray transient \citep{Wijnands2006}.
 
The intensity, amplitude, duration and X-ray spectral properties of the outburst of \igr\ strongly suggest that this new X-ray transient is an LMXB. While the \swift\ spectra are well described by a simple absorbed powerlaw model with a photon index  $\Gamma$ of $\sim$2, the \xmm\ spectra need an additional thermal component to be adequately modelled. This thermal component could possibly arise from an accretion disc (a disc blackbody model can adequately fit the data), in which case we cannot determine whether the accreting object is a neutron star or a black hole. However, accretion discs at these low luminosities are typically expected to be at lower temperatures than what we observe, which suggests that the thermal component may have a different origin. In fact, the presence of the soft component (and its associated temperature) in the X-ray spectrum of \igr\ is reminiscent of what is seen in neutron star X-ray binaries at similar luminosities \citep[e.g.,][]{Fridriksson2010,ArmasPadilla2013b,Degenaar2013}. In those systems the thermal component it thought to originate from the surface of the neutron star. Given the strong similarities, we tentatively identify the accretor in \igr\ as a neutron star.

If the soft component is indeed originating from the neutron star surface, then it is likely due to low level accretion onto
the surface. Such low level accretion will indeed produce a soft
spectrum \citep{Zampieri1995} and in principle
one should use the so-called zamp model \citep{Soria2011} to fit the soft component in the X-ray spectra. However, that model is not
publicly available so the next best things are the neutron star
atmosphere models. Although those models assume that the emission is
from a cooling neutron star and therefore incorporate different
microphysics compare to the zamp model, typically the results are
very similar between the models (i.e., the inferred surface
temperature; \citealt{Soria2011}). Therefore, we
will discuss the results obtained using neutron star atmosphere models
which also allows for direct comparison with previous work in which
only neutron star atmosphere models were used.

During the \xmm\ observation we see a steady decrease of the source
intensity. Together with the intensity, the temperature of the soft
component is decreasing as well. Although such a temperature decrease
with decreasing intensity would be expected irrespectively if the soft
component is due to a cooling accreting disc or due to decreasing
accretion onto the neutron star surface, it is very similar to what
recently has been found for the bright neutron-star X-ray transient \xtecool\
by \citet{Degenaar2013}. Those authors found that a similar (albeit at
slightly lower luminosities and lower inferred surface temperatures)
decay during the end stages of the 2012 outburst of \xtecool. They
favored an interpretation that the observed temperature decrease was
due to the cooling of the neutron star crust which was heated by the
accretion during the outburst.  

However, the outburst of \igr\ was much shorter (3-4 weeks versus
10 weeks) and much less luminous (peaking at $\sim 7\times10^{35}$ erg
s$^{-1}$ versus $\sim5 \times 10^{37}$ erg s$^{-1}$; 2--10 keV)
than that of \xtecool. Within our understanding of crustal
heating and cooling models, \igr\ should have been heated to a much
lesser extent (if at all). It would not be expected to display
similarly strong signs of crustal cooling as \xtecool. Therefore, this
likely is not the correct interpretation of the observed rapid
intensity decay during the \xmm\ observation for \igr. Since the behaviour of this source is so similar as what has been observed for \xtecool, likely also in that source we did not observe the crust cooling.

\citet{Degenaar2013} alternatively suggested that the
temperature decrease observed for \xtecool\ was not due to a cooling
crust but maybe the thermal component was due to low-level accretion on
the neutron star surface and the accretion rate decreased
causing the surface temperature to decrease as well.  This scenario
could more naturally account for the similarities in the decay seen
for the two sources, despite their different outburst
properties. Therefore, we favor a low-level accretion scenario to
explain the thermal component in both sources. If this is
indeed the correct interpretation for the decay seen in both sources,
then there is no need for an additional heat source in the neutron
star crust, as was proposed by \citet{Degenaar2013}.

Moreover, investigating the literature for more detections of soft
thermal components at such low accretion luminosities of accreting
neutron star systems (or candidate neutron star systems), we find
several additional sources which are consistent with the hypothesis
that the soft thermal component in the luminosity range of $10^{34}$
to $10^{35}$ erg s$^{-1}$ is due to low level accretion onto the
neutron star surface. For example, for the suspected neutron star
very-faint X-ray transient \xte\ and for two confirmed persistent
 neutron star VFXBs (\rxh, \rxs) we found in
previous work that their spectra also required a thermal component in
addition to the power-law component
\citep{ArmasPadilla2011,ArmasPadilla2013b}.  We refitted their spectra with
the NSATMOS model (we previously used blackbody models) and obtained temperatures of $0.162\pm0.004 $~keV, $0.184\pm0.004
$~keV and $0.194\pm0.003$~keV, respectively, with associated luminosities of
3.3$\times10^{34}$, $4.4\times10^{34}$, and $6.7\times10^{34}$ erg
s$^{-1}$. In addition, \citet{Fridriksson2010} reported on a
brief accretion flare during the quiescent state of the
neutron-star X-ray transient \xtejoel\ during which the temperature
was 0.159$\pm$0.02 keV for a luminosity of $2.6\times 10^{34}$ erg s$^{-1}$.

Uncertainties in the distances affect the inferred temperatures, but
there appears to be a tendency that for higher luminosities,
the temperature is higher, as one would expect for the low-level
accretion scenario. This provides strong evidence that in all these
sources the thermal component indeed arises from the neutron star
surface as a result of low-level accretion and that the inferred
temperature is determined by the instantaneous mass accretion rate.

The decrease of surface temperature while the luminosity (and thus the
inferred accretion rate onto the neutron star surface)
decreases can only continue (in the transient sources) as long as the
temperature due to the low-level accretion is higher than the interior
(crust) temperature of the neutron star. If the accretion rate drops
below a certain value, then the light curve evolution will not
be governed anymore by the decay in accretion rate but instead it will
be determined by how fast the crust cools down (even small outbursts
will have a slightly heated crust) and eventually by the core cooling
rate. Therefore, in the light curve we would expect a break at a
certain luminosity from a rapid decay of the luminosity to a
much slower decay rate. Such a break has been observed in several
systems (e.g., \xtejoel\ by \citealt{Fridriksson2010}; MAXI J0556--332 by Homan et al. 2013, in
prep; Aql X-1; \citealt{Campana1998}; Campana et
al. 2013, in prep) and indeed has been interpreted as the onset of the crust
cooling \citep[e.g., ][]{Fridriksson2010}.

\vspace{-0.6cm}
\section*{Acknowledgments}

\vspace{-0.2cm}
We acknowledge the \xmm\ team for make this observation possible. RW
and MAP are supported by an ERC starting grant
awarded to RW. ND is supported by a NASA Hubble Postdoctoral
Fellowship grant (number HST-HF-51287.01-A) from the Space Telescope
Science Institute.

\vspace{-0.4cm}


\label{lastpage}
\end{document}